\documentclass[12pt]{article}
\usepackage[latin9]{inputenc}
\usepackage{amsmath}
\usepackage[unicode=true,pdfusetitle,
 bookmarks=true,bookmarksnumbered=false,bookmarksopen=false,
 breaklinks=false,pdfborder={0 0 1},backref=false,colorlinks=false]
 {hyperref}

\makeatletter

\DeclareTextSymbolDefault{\textquotedbl}{T1}

\@ifundefined{date}{}{\date{}}
\usepackage{tikz}
 \usepackage{stmaryrd}
  \usepackage{stackrel}

\usepackage{latexsym}
\newcommand{\be}{\begin{equation}}
\newcommand{\ee}{\end{equation}}
\newcommand{\bea}{\begin{eqnarray}}
\newcommand{\eea}{\end{eqnarray}}
\newcommand{\beas}{\begin{eqnarray*}}
\newcommand{\eeas}{\end{eqnarray*}}

\def\tr{{\rm Tr}}

\def\XXint#1#2#3{{\setbox0=\hbox{$#1{#2#3}{\int}$ }
\vcenter{\hbox{$#2#3$ }}\kern-.5\wd0}}

\makeatother

\begin{document}
\title{Constructing the bulk at the critical point of three-dimensional large
$N$ vector theories}
\author{Celeste Johnson\thanks{Email: cel.284@gmail.com} $^{a}$ , 
Mbavhalelo Mulokwe\thanks{Email: mulokwe.mbavhalelo@gmail.com} $^{b}$
and Jo\~ao P. Rodrigues\thanks{Email: joao.rodrigues@wits.ac.za} $^{a}$
\\
 \\
 $^{a}$National Institute for Theoretical Physics \\
 School of Physics and Mandelstam Institute for Theoretical Physics
\\
 University of the Witwatersrand, Johannesburg\\
 Wits 2050, South Africa \\
 \\
 $^{b}$ Department of Physics, \\
 University of Pretoria \\
 Private Bag X20, Hatfield 0028, South Africa \\
 }
\maketitle
\begin{abstract}
In the context of the $AdS_{4}/CFT_{3}$ correspondence between higher
spin fields and vector theories, we use the constructive bilocal
fields based approach to this correspondence, to demonstrate, at the
$IR$ critical point of the interacting vector theory and directly
in the bulk, the removal of the $\Delta=1$ ($s=0$) state from the
higher spins field spectrum, and to exhibit simple Klein-Gordon higher
spin Hamiltonians. The bulk variables and higher spin fields are obtained
in a simple manner from boundary bilocals, by the change of variables
previously derived for the $UV$ critical point (in momentum space),
together with a field redefinition. 
\end{abstract}

\section{Introduction}

The AdS/CFT correspondence \cite{Maldacena:1997re,Gubser:1998bc,Witten:1998qj}
has a very interesting application in the context of the higher spin
theories/vector model correspondence \cite{Klebanov:2002ja}. Of particular
interest to us is the $AdS_{4}/CFT_{3}$ correspondence \footnote{There is a vast literature on the subject; \cite{Fronsdal:1978rb}
- \cite{Giombi:2013fka} are representative of the work on the subject,
but they do not form by any means an exhaustive list.}. Although the higher spin degrees of freedom of Fronsdal and Vasiliev
are not those of string theory\footnote{ For attempts to link the two, see for instance \cite{Chang:2012kt}
- \cite{Honda:2017nku}}, there are several reasons why this correspondence is of importance
and deserves further study. These include the absence of supersymmetry
and the fact that vector models are \textquotedbl solvable\textquotedbl{}
in the large $N$ limit, allowing for a more concrete and detailed
study of the workings of the correspondence, and possibly even providing
a definition of (gauge fixed) higher spin theories themselves, through
their dual vector valued field theories.

We focus in this communication on the constructive approach of \cite{Das:2003vw,Koch:2010cy,Jevicki:2011ss,deMelloKoch:2012vc,Koch:2014aqa}.
In this approach, the singlet sector of $O(N)$ invariant field theories
is described in terms of equal time bilocals, appropriate to an Hamiltonian
description of the theory,

\begin{equation}
\psi_{\vec{x}_{1}\vec{x}_{2}}=\sum_{a=1}^{N}\phi^{a}\left(t,\vec{x}_{1}\right)\phi^{a}\left(t,\vec{x}_{2}\right),
\end{equation}

\noindent where $\vec{x}_{1}$ and $\vec{x}_{2}$ are two dimensional
space vectors. For the free theory (the $UV$ fixed point), these
5 degrees of freedom and their canonical conjugates are mapped to
$AdS_{4}\times S_{1}$, where the $S_{1}$ encodes the spin degrees
of freedom. The map is a phase transformation, but is a point transformation
in momentum space. In a temporal gauge\cite{Koch:2014aqa}, it is
given by:

\begin{align}
E & =E_{1}+E_{2}=|\vec{p_{1}}|+\vec{p_{2}}|\label{free energies}\\
\vec{p} & =\vec{p_{1}}+\vec{p}_{2}\label{map:2}\\
p^{z} & =2\sqrt{\left|\vec{p_{1}}\right|\left|\vec{p_{2}}\right|}\sin\left(\frac{\varphi_{2}-\varphi_{1}}{2}\right)\label{map:3}\\
\theta & =\arctan\left(\frac{2\vec{p}_{2}\times\vec{p}_{1}}{\left(\left|\vec{p_{1}}\right|-\left|\vec{p_{2}}\right|\right)p^{z}}\right)\label{map:4}
\end{align}
with $\varphi_{2}-\varphi_{1}$ being the angle between $\vec{p_{1}}$
and $\vec{p_{2}}$, and $\vec{p}_{2}\times\vec{p}_{1}\equiv p_{2}^{1}p_{1}^{2}-p_{2}^{2}p_{1}^{1}$
\cite{Koch:2014aqa}.

The three dimensional $O(N)$ vector theory with a $\frac{\lambda}{N}(\phi^{a}\phi^{a})^{2}$
interaction has an IR fixed critical point. At this critical point,
the theory is expected to contain a state with dimension $\Delta=2$,
a boundary field in the standard AdS/CFT correspondence with the standard
positive branch for the expression of the dimension of the operator
\cite{Klebanov:2002ja}, and no longer the $\Delta=1$ state present
in the UV critical point. Although general arguments exist relating
the two through a Legendre transformation \cite{Petkou:2003zz}, in
practice the IR fixed point is described in terms of a non-linear
sigma model \cite{ZinnJustin,Lang:1990ni}. In this description, the
Lagrange multiplier field is naturally identified with the $\Delta=2$
state, but it is certainly not apparent that the $\Delta=1$ is no
longer present in the theory, or equivalently, that the constraint
is enforced beyond the leading large N order.


These issues were discussed and successfully resolved in \cite{Mulokwe:2018czu}
directly in terms of the $\frac{\lambda}{N}(\phi^{a}\phi^{a})^{2}$
theory, using bilocal fields on the field theory boundary. The two
point function for bilocals appropriate to the path integral description
of the boundary field theory (a Bethe-Sapeter equation \cite{deMelloKoch:1996mj}
in terms of the original field theory variables) was obtained, and
shown to take a universal form at the $IR$ critical point\footnote{Path integral bilocal holography was previously discussed in \cite{deMelloKoch:2014mos},
and more recently with the use of conformal group techniques in \cite{deMelloKoch:2018ivk},
\cite{Aharony:2020omh}. }. It consists of the two free propagators present in the $UV$ limit
plus a connected piece with a pole identified with the $\Delta=2$
state. This bilocal propagator was then shown to be equivalent to
the spectrum equation arising in the Hamiltonian bilocal approach
as a result of integration of an intermediate energy variable. In
both cases, the absence of a boundary $\Delta=1$ state was demonstrated.

In this communication, we address the question of whether the map
(\ref{free energies}) - (\ref{map:4}) and the construction of bulk
fields, established for the $UV$ critical point, is still applicable
at the $IR$ fixed point, or if it needs adjusting. It will be shown
that the map remains valid, and that by introducing a suitable field
redefinition in the definition of bulk higher spin fields, the connected
piece of the propagator / spectrum equation precisely removes the
$s=0$ state from the bulk higher spin field. 

This letter is organised as follows. Overall, Section $2$ discusses
the bilocal description of the boundary. Subsection $2.1$ briefly
describes the conformal IR fixed point of the $\frac{\lambda}{N}(\phi^{a}\phi^{a})^{2}$
theory at leading order in $N$. In Subsection 2.2, the bilocal spectrum
equation of the $1/N$ quadratic Hamiltonian fluctuations is obtained,
and a potential scattering problem ensues. In Subsection $2.3$,
the most general solution to the spectrum equation is obtained, and
is shown to take a universal form at IR criticality. Section $3$
describes the construction of the bulk. Using a change of variables
from bilocal momenta to bulk momenta (as dictated by the map (\ref{free energies})
- (\ref{map:4})), and instituting a field redefinition to define
bulk higher spin fields, a bulk Hamiltonian is obtained in Subsection
$3.1$ for the free case that is simply the sum of Hamiltonians of
massless spin $s$ fields in an equal time slice of $AdS_{4}$. In
Subsection $3.2$, again using the same field redefinition and the
map (\ref{free energies}) - (\ref{map:4}), we are able to obtain
the bulk description of the universal boundary eigenstates at the
IR critical point. It is then shown directly in the bulk that the
$s=0$ (or $\Delta=1$ state) is exactly removed from the spectrum.
It is remarkable that this direct construction of the bulk is obtained
by a simple change of variables (\ref{free energies}) - (\ref{map:4})
accompanied by a field redefinition in defining bulk higher spin fields
from boundary bilocals. Subsection $3.3$ exhibits how the bulk $\Delta=2$
state becomes a boundary state at IR criticality. In addition, we show explicitly that the bulk $AdS_{4}\times S_{1}$
Hamiltonian projects to the boundary Hamiltonian with the correct
dispersion relation for a single mode bound state. This was the expected
result and serves as a further check of our bulk higher spin field redefinitions. Section $4$ is
left for a brief discussion and outlook.



\section{Bilocal boundary}

\subsection{Bilocal Hamiltonian and large-$N$ conformal background}

Our starting point is the Hamiltonian density of a three (space-time)
dimensional scalar vector theory with a quartic interaction $\frac{\lambda}{N}(\phi^{a}\phi^{a})^{2},a=1,...,N$:
\begin{eqnarray*}
\mathcal{H}=\frac{1}{2}\pi^{a}\pi^{a}+\frac{1}{2}\vec{\nabla}\phi^{a}\cdot\vec{\nabla}\phi^{a}+\frac{1}{2}m^{2}\phi^{a}\phi^{a}+\frac{\lambda}{4!N}\left(\phi^{a}\phi^{a}\right)^{2},\,\pi^{a}(\vec{x})=-i\frac{\partial}{\partial\phi^{a}(\vec{x})}.
\end{eqnarray*}
We use the collective field theory method \cite{Jevicki:1979mb} to
re-express the above Hamiltonian in terms of $O(N)$ invariant equal
time bilocals 
\begin{eqnarray}
\psi_{\vec{x}_{1}\vec{x}_{2}}(t)=\sum_{a=1}^{N}\phi^{a}\left(t,\vec{x}_{1}\right)\phi^{a}\left(t,\vec{x}_{2}\right),
\end{eqnarray}
and their canonical conjugates, as appropriate to an Hamiltonian approach.
This is achieved by a simple change of variables from the original
fields of the scalar theory to the invariant bilocals, and by a similarity
transformation: 
\begin{eqnarray*}
\partial_{\alpha}\rightarrow\partial_{\alpha}-\frac{1}{2}\partial_{\alpha}\ln J\,,\,\alpha\equiv\psi_{\vec{x_{1}}\vec{x_{2}}}.
\end{eqnarray*}
$J$ is the Jacobian induced by the change of variables, and the above
transformation ensures that the collective Hamiltonian is explicitly
hermitian. For vector models the large $N$ form of the Jacobian is
known (see for instance \cite{Rodrigues:1992ru}, \cite{deMelloKoch:1996mj})
and its leading large $N$ form is given by: 
\[
\ln J=\frac{N}{2}\tr\ln\psi.
\]
The trace is in (spatial) functional space. One obtains the form of
the collective field theory Hamiltonian sufficient to generate the
large $N$ background and spectrum: 
\begin{eqnarray}
H & = & \frac{2}{N}\mathrm{Tr}\Pi\psi\Pi+\frac{N}{8}\mathrm{Tr}(\psi^{-1})\nonumber \\
 &  & +N\int d^{d-1}\vec{x}\left(-\frac{1}{2}\lim_{\vec{x}\rightarrow\vec{y}}\partial_{\vec{y}}^{2}\psi_{\vec{x}\vec{y}}+\frac{1}{2}m^{2}\psi_{\vec{x}\vec{x}}+\frac{\lambda}{4!}\psi_{\vec{x}\vec{x}}^{2}\right)\nonumber \\
 & \equiv & \frac{2}{N}\mathrm{Tr}\Pi\psi\Pi+NV_{eff},\hspace{15pt}\Pi_{\vec{x}\vec{y}}=-i\frac{\partial}{\partial\psi_{\vec{x}\vec{y}}}.
\end{eqnarray}
The fields have been rescaled $\psi\to N\psi$ to make explicit the
$N$ dependence. In the large $N$ limit the kinetic term is subleading,
and with the large $N$ translationally invariant ansatz: 
\begin{equation}
\psi_{\vec{x}\vec{y}}^{0}=\int\frac{d^{2}k}{\left(2\pi\right)^{2}}e^{i\vec{k}\cdot\left(\vec{x}-\vec{y}\right)}\psi_{\vec{k}}^{0}\,,
\end{equation}
the standard gap equation 
\begin{equation}
s=\frac{1}{2}\int\frac{d^{2}\vec{k}}{\left(2\pi\right)^{2}}\frac{1}{\sqrt{\vec{k}^{2}+m^{2}+\frac{\lambda}{6}s}},\hspace{10pt}s=\int\frac{d^{2}\vec{k}}{\left(2\pi\right)^{2}}\psi_{\vec{k}}^{0}.
\end{equation}
is obtained. Defining $\alpha\equiv m^{2}+\frac{\lambda}{6}s$, one
has\footnote{Our notations is as follows: $k=(E,\vec{k})$ with Minkowski signature
$(+,-,-)$ and $k_{E}$ is the euclidean momentum $3$-vector. } 
\[
\frac{6}{\lambda}(\alpha-m^{2})=\int\frac{d^{2}\vec{k}}{\left(2\pi\right)^{2}}\frac{1}{2\sqrt{\vec{k}^{2}+\alpha}}=\int\frac{d^{3}k}{\left(2\pi\right)^{3}}\frac{i}{k^{2}-\alpha}=\int\frac{d^{3}k_{E}}{\left(2\pi\right)^{3}}\frac{1}{k_{E}^{2}+\alpha}.
\]
Our regularization is defined as: 
\begin{equation}
\int\frac{d^{d}k_{E}}{\left(2\pi\right)^{d}}\frac{1}{k_{E}^{2}+\alpha}=\frac{1}{\left(4\pi\right)^{d/2}}\Gamma\left(1-\frac{d}{2}\right)\alpha^{\frac{d-2}{2}}\to-\frac{1}{4\pi}\sqrt{\alpha},\,\text{for}\,\,d=3.\label{regularization}
\end{equation}
Thus one obtains the equation $\alpha+\frac{\lambda}{24\pi}\sqrt{\alpha}-m^{2}=0$.
The IR fixed point is associated with the root: 
\begin{equation}
\sqrt{\alpha}=\frac{24\pi m^{2}}{\lambda}+O(\frac{m^{4}}{\lambda^{3}})\label{mass}
\end{equation}

\noindent and is approached by keeping $m^{2}$ finite and taking
$|\lambda|\to\infty$. At the critical point then, the background
propagator takes the conformal form:

\begin{equation}
\psi_{\vec{k}}^{0}=\frac{1}{2|\vec{k}|}.
\end{equation}

\noindent and is the $O(N)$ invariant two point function of the underlying
scalar fields.

\subsection{Quadratic Hamiltonian and spectrum equation}

$1/N$ corrections yield the spectrum, which is obtained from small
fluctuations about the large-N conformal background. One shifts, 
\begin{eqnarray*}
\psi_{\vec{x}_{1}\vec{x}_{2}} & = & \psi_{\vec{x}_{1}\vec{x}_{2}}^{0}+\frac{1}{\sqrt{N}}\eta_{\vec{x}_{1}\vec{x}_{2}};\hspace{0.7cm}\Pi_{\vec{x}_{1}\vec{x}_{2}}=\sqrt{N}\pi_{\vec{x}_{1}\vec{x}_{2}},
\end{eqnarray*}
from which the quadratic Hamiltonian follows: 
\begin{equation}
H^{(2)}=2\mathrm{Tr}\left(\pi\psi^{0}\pi\right)+\frac{1}{8}\mathrm{Tr}\left((\psi^{0})^{-1}\eta(\psi^{0})^{-1}\eta(\psi^{0})^{-1}\right)+\frac{\lambda}{4!}\int d^{2}\vec{x}\eta_{\vec{x}\vec{x}}^{2}.\label{quadham}
\end{equation}
The equations of motion for $\eta$ are then: 
\begin{align*}
{\ddot{\eta}}_{\vec{x_{1}}\vec{x_{2}}} & =%
-\frac{1}{4}\left[{(\psi^{0})}^{-1}\eta(\psi^{0})^{-1}+\eta(\psi^{0})^{-2}+(\psi^{0})^{-2}\eta+(\psi^{0})^{-1}\eta(\psi^{0})^{-1}\right]_{\vec{x_{1}}\vec{x_{2}}}\\
 & -\frac{\lambda}{6}\left(\psi_{\vec{x_{1}}\vec{x_{2}}}^{0}(\eta_{\vec{x_{1}}\vec{x_{1}}}+\eta_{\vec{x_{2}}\vec{x_{2}}})\right)%
.
\end{align*}
Looking for eigen-frequencies, and Fourier transforming: 
\[
\eta_{\vec{x_{1}}\vec{x_{2}}}(t)=e^{-iEt}\eta_{\vec{x_{1}}\vec{x_{2}}}\,,\hspace{10pt}\eta_{\vec{x_{1}}\vec{x_{2}}}=\int\frac{d^{2}\vec{k}_{1}}{2\pi}\int\frac{d^{2}\vec{k}_{2}}{2\pi}e^{i\vec{k}_{1}\vec{x}_{1}+i\vec{k}_{2}\vec{x}_{2}}\eta_{\vec{k}_{1}\vec{k}_{2}}\,,
\]
one obtains the spectrum equation: 
\begin{equation}
E^{2}\eta_{\vec{k}_{1}\vec{k}_{2}}=\frac{1}{4}\left((\psi_{\vec{k}_{1}}^{0})^{-1}+(\psi_{\vec{k}_{2}}^{0})^{-1}\right)^{2}\eta_{\vec{k}_{1}\vec{k}_{2}}+\frac{\lambda}{6}\left(\psi_{\vec{k}_{1}}^{0}+\psi_{\vec{k}_{2}}^{0}\right)\int\frac{d^{2}\vec{l}}{(2\pi)^{2}}\eta_{\vec{k}_{1}+\vec{k}_{2}-\vec{l},\vec{l}}.\label{eval_eqn}
\end{equation}
At the UV point ($\lambda=0$), the large $N$ background is also
conformal, and 
\begin{equation}
E_{\vec{k}_{1}\vec{k}_{2}}^{2}=\frac{1}{4}\left((\psi_{\vec{k_{1}}}^{0})^{-1}+(\psi_{\vec{k_{2}}}^{0})^{-1}\right)^{2}=\left(|\vec{k_{1}}|+|\vec{k_{2}}|\right)^{2},\label{free}
\end{equation}
a result known for some time \cite{Jevicki:1983hb} and at the root
of the $AdS_{4}/CFT_{3}$ constructive map \cite{Koch:2010cy,Koch:2014aqa}
at the free UV fixed point. At the IR fixed point, the spectrum is
to be understood as that of a quantum mechanical (relativistic) potential
scattering problem for the set of continuum states with $E_{\vec{k}_{1}\vec{k}_{2}}^{2}=\left(|\vec{k_{1}}|+|\vec{k_{2}}|\right)^{2}.$
It can then be expected that the $AdS_{4}/CFT_{3}$ constructive map
of \cite{Koch:2014aqa} remains valid.

\subsection{States on the bilocal boundary}

As is well known, the most general solution of the spectrum equation
(\ref{eval_eqn}) for potential scattering with (squared) energy $E_{\vec{p}_{1}\vec{p}_{2}}^{2}=(|\vec{p_{1}}|+|\vec{p}_{2}|)^{2}$
can be written as: 
\[
\eta_{\vec{k_{1}},\vec{k_{2}}}^{\vec{p}_{1},\vec{p}_{2}}=\rho_{\vec{k_{1}},\vec{k_{2}}}^{\vec{p}_{1}\vec{p}_{2}}+\frac{\lambda}{12}\frac{1}{E_{\vec{p}_{1}\vec{p}_{2}}^{2}-(|\vec{k_{1}}|+|\vec{k_{2}}|)^{2}}\Big(\frac{1}{|\vec{k_{1}}|}+\frac{1}{|\vec{k_{2}}|}\Big)\int\frac{d^{2}\vec{l}}{(2\pi)^{2}}\eta_{\vec{k_{1}}+\vec{k_{2}}-\vec{l},\vec{l}}^{\vec{p}_{1}\vec{p}_{2}}\,\,,
\]

\noindent where $\rho_{\vec{k_{1}},\vec{k_{2}}}^{\vec{p}_{1}\vec{p}_{2}}$
is a solution of the free equation, which we normalize to $\rho_{\vec{k_{1}},\vec{k_{2}}}^{\vec{p}_{1}\vec{p}_{2}}=\delta^{2}(\vec{p_{1}}-\vec{k_{1}})\delta^{2}(\vec{p_{2}}-\vec{k_{2}})$.
In the above, $(\vec{p_{1}},\vec{p_{2}})$ labels the states and $(\vec{k_{1}},\vec{k_{2}})$
are momentum coordinates.

Integration of both sides of the full scattering solution results
in (\cite{Mulokwe:2018czu}) 
\begin{eqnarray}
\int\frac{d^{2}l}{(2\pi)^{2}}\eta_{\vec{k_{1}}+\vec{k_{2}}-\vec{l},\vec{l}}^{\vec{p}_{1},\vec{p}_{2}} & = & \frac{\delta^{2}(\vec{p}_{1}+\vec{p}_{2}-\vec{k}_{1}-\vec{k}_{2})}{(2\pi)^{2}\left(1+\frac{\lambda}{6i}\int\frac{d^{3}l}{(2\pi)^{3}}\frac{1}{l^{2}(p_{1}+p_{2}-l)^{2}}\right)},\label{bum}
\end{eqnarray}
where we have used the result \cite{Mulokwe:2018czu} 
\begin{equation}
\frac{1}{E_{\vec{p}}^{2}-(|\vec{l}|+|\vec{p}-\vec{l}|)^{2}}\Big(\frac{1}{|\vec{l}|}+\frac{1}{|\vec{p}-\vec{l}|}\Big)=2i\int\frac{dE_{l}}{(2\pi)}\frac{1}{l^{2}(p-l)^{2}}.
\end{equation}
Since 
\begin{eqnarray*}
\int\frac{d^{3}l}{(2\pi)^{3}}\frac{1}{l^{2}(p-l)^{2}} & = & \frac{i}{8\left|p_{E}\right|},
\end{eqnarray*}
the form of the scattering solution for finite $\lambda$ is: 
\begin{align}
\eta_{\vec{k}_{1},\vec{k}_{2}}^{\vec{p}_{1},\vec{p}_{2}} & =\delta^{2}(\vec{p}_{1}-\vec{k}_{1})\delta^{2}(\vec{p}_{2}-\vec{k}_{2})\label{finite_lambda}\\
 & +\frac{\delta^{2}(\vec{p}_{1}+\vec{p}_{2}-\vec{k}_{1}-\vec{k}_{2})}{E_{\vec{p}_{1}\vec{p}_{2}}^{2}-(|\vec{k}_{1}|+|\vec{k}_{2}|)^{2}}\Big(\frac{1}{|\vec{k}_{1}|}+\frac{1}{|\vec{k}_{2}|}\Big)\Big(\frac{\lambda}{48\pi^{2}}\Big)\frac{1}{1+\frac{\lambda}{48\left|p_{E}\right|}},\nonumber 
\end{align}
with $|{p}_{E}|=\sqrt{-\left(|p_{1}|+|p_{2}|\right)^{2}+\left(\vec{p}_{1}+\vec{p}_{2}\right)^{2}}$.
At the IR critical point ($|\lambda|\rightarrow\infty$), the scattering
states take a universal critical form: 
\begin{align}
\eta_{\vec{k}_{1},\vec{k}_{2}}^{\vec{p}_{1},\vec{p}_{2}} & =\delta^{2}(\vec{p}_{1}-\vec{k}_{1})\delta^{2}(\vec{p}_{2}-\vec{k}_{2})\nonumber \\
 & +\frac{\left|p_{E}\right|}{\pi^{2}}\frac{\delta^{2}(\vec{p}_{1}+\vec{p}_{2}-\vec{k}_{1}-\vec{k}_{2})}{E_{\vec{p}_{1}\vec{p}_{2}}^{2}-\left(|\vec{k}_{1}|+|\vec{k}_{2}|\right)^{2}}\left(\frac{1}{|\vec{k}_{1}|}+\frac{1}{|\vec{k}_{2}|}\right)\label{finite_scattering}
\end{align}
where $E_{\vec{p}_{1}\vec{p}_{2}}^{2}=\left(|\vec{p}_{1}|+|\vec{p}_{2}|\right)^{2}$.
On the boundary, that the $\Delta=1$ state is no longer in the spectrum
is more simply shown by taking the limit $|\lambda|\to\infty$ in
equation (\ref{bum}) \footnote{$\eta_{\vec{x}\vec{x}}=\int d^{2}pe^{i\vec{p}\cdot\vec{x}}\int\frac{d^{2}l}{(2\pi)^{2}}\eta_{\vec{p}-\vec{l},\vec{l}}$
is a boundary field, as $z\sim(\vec{x}_{1}-\vec{x}_{2})\cdot\vec{f}(\vec{p}_{1},\vec{p}_{2})$}. Alternatively it can also be confirmed directly from the critical
form (\ref{finite_scattering}), by integration with $\vec{k_{1}}+\vec{k_{2}}$
fixed, and using the integral results stated in the above. This agrees
with results obtained with path integral bilocal correlators at criticality
\cite{Mulokwe:2018czu}. 

Bound states are well known to correspond to eigenspectrum solutions
in the absence of an incident wave, or equivalently as particular
solutions of (\ref{eval_eqn}): 
\[
\eta_{\vec{k}_{1}\vec{k}_{2}}^{B}=\frac{\lambda}{12}\frac{1}{E^{2}-\left(|\vec{k}_{1}|+|\vec{k}_{2}|\right)^{2}}\left(\frac{1}{|\vec{k}_{1}|}+\frac{1}{|\vec{k}_{2}|}\right)\int\frac{d^{2}l}{(2\pi)^{2}}\eta_{\vec{k_{1}}+\vec{k_{2}}-\vec{l},\vec{l}}^{B}
\]
Use of the integral results stated above determines its energy to
be 
\[
E^{2}=(\vec{k}_{1}+\vec{k}_{2})^{2}-(\frac{\lambda}{48})^{2}
\]
As expected, bound states can also be identified as poles in the connected
piece (transmission amplitude) of (\ref{finite_lambda}) ocuring at
$48|p_{E}|=-\lambda$.


This is a state of infinite (tachyon) squared mass present as $\lambda\to-\infty$.
It appears as an infinite pole $|p_{E}|\to\infty$ in the universal
connected piece of (\ref{finite_scattering}). This is the $\Delta=2$
state \cite{Mulokwe:2018czu}\footnote{In $3$ euclidean dimensions $\int\frac{d^{3}x}{x^{4}}e^{ik_{E}x}\sim|k_{E}|$}.

\section{Constructing the bulk}

\subsection{Higher spin fields in the bulk - field redefinition and quadratic
Hamiltonian}

We now wish to use the map (\ref{free energies}) - (\ref{map:4})
to explicitly construct the higher spin fields in the bulk. We first
discuss the free case.

The Jacobian for the change of variables from bilocals to $AdS$ coordinates
\cite{Koch:2014aqa} is given by 
\begin{eqnarray}
\left|\frac{\partial\vec{k}_{AdS\times S^{1}}}{\partial\vec{k}_{\mathrm{bilocal}}}\right|=\frac{1}{|\vec{k}_{1}|}+\frac{1}{|\vec{k}_{2}|}.\label{B}
\end{eqnarray}
We use the notation $\vec{k}_{AdS\times S^{1}}=(\vec{k},k^{z},\theta)\equiv\vec{\kappa}$
and $\vec{k}_{\mathrm{bilocal}}=(\vec{k}_{1},\vec{k}_{2})$. The equal
time slice is the same.

We wish to preserve the canonical structure under this change of variables:
\begin{eqnarray*}
\left[\pi_{\vec{k}_{1}\vec{k}_{2}},\eta_{\vec{k}'_{1}\vec{k}'_{2}}\right] & = & -i\delta^{2}(\vec{k}_{1}-\vec{k}'_{1})\delta^{2}(\vec{k}_{2}-\vec{k}'_{2})\\
 & = & -i\left(\frac{1}{|\vec{k}_{1}|}+\frac{1}{|\vec{k}_{2}|}\right)\delta(\vec{k}_{AdS\times S^{1}}-\vec{k}'_{AdS\times S^{1}})\\
\Rightarrow\frac{\left[\pi_{\vec{k}_{1}\vec{k}_{2}},\eta_{\vec{k}'_{1}\vec{k}'_{2}}\right]}{\frac{1}{|\vec{k}_{1}|}+\frac{1}{|\vec{k}_{2}|}} & = & -i\delta(\vec{k}_{AdS\times S^{1}}-\vec{k}'_{AdS\times S^{1}})
\end{eqnarray*}
This requries a redefinition of at least one of the fields.

Let us now consider the form of the (free) quadratic Hamiltonian (\ref{quadham})
in momentum space: 
\begin{eqnarray*}
H_{2} & = & \int d^{2}{k}_{1}\int d^{2}{k}_{2}\hspace{3pt}\left(\pi_{\vec{k}_{1}\vec{k}_{2}}\left(\psi_{\vec{k}_{1}}^{0}+\psi_{\vec{k}_{2}}^{0}\right)\pi_{-\vec{k}_{2}-\vec{k}_{1}}\right)\\
 &  & +\frac{1}{16}\int d^{2}{k}_{1}\int d^{2}{k}_{2}\hspace{3pt}\eta_{\vec{k}_{1}\vec{k}_{2}}\\
 &  & \left((\psi_{\vec{k}_{1}}^{0})^{-2}(\psi_{\vec{k}_{2}}^{0})^{-1}+(\psi_{\vec{k}_{2}}^{0})^{-2}(\psi_{\vec{k}_{1}}^{0})^{-1}\right)\eta_{-\vec{k}_{2}-\vec{k}_{1}}\\
 & = & \int d^{2}{k}_{1}\int d^{2}{k}_{2}\hspace{3pt}\frac{1}{2}\left(\pi_{\vec{k}_{1}\vec{k}_{2}}\left(\frac{1}{|\vec{k}_{1}|}+\frac{1}{|\vec{k}_{2}|}\right)\pi_{-\vec{k}_{2}-\vec{k}_{1}}\right)\\
 &  & +\frac{1}{2}\eta_{\vec{k}_{1}\vec{k}_{2}}\left(|\vec{k}_{1}|^{2}|\vec{k}_{2}|+|\vec{k}_{2}|^{2}|\vec{k}_{1}|\right)\eta_{-\vec{k}_{2}-\vec{k}_{1}}.
\end{eqnarray*}
By factorizing the Jacobian, this can be re-written as 
\begin{eqnarray}
H_{2} & = & \int d^{2}{k}_{1}\int d^{2}{k}_{2}\hspace{3pt}\left(\frac{1}{|\vec{k}_{1}|}+\frac{1}{|\vec{k}_{2}|}\right)\frac{1}{2} \left[    \left(\pi_{\vec{k}_{1}\vec{k}_{2}}\pi_{-\vec{k}_{2}-\vec{k}_{1}}\right)\right.\nonumber \\
 &  & \left.+\frac{\eta_{\vec{k}_{1}\vec{k}_{2}}\left(|\vec{k}_{1}|+|\vec{k}_{2}|\right)^{2}\eta_{-\vec{k}_{2}-\vec{k}_{1}}}{\left(\frac{1}{|\vec{k}_{1}|}+\frac{1}{|\vec{k}_{2}|}\right)\left(\frac{1}{|\vec{k}_{1}|}+\frac{1}{|\vec{k}_{2}|}\right)}\right].\label{HS_ham1}
\end{eqnarray}
This then suggests that we define the bulk higher spin field and its
conjugate field as \cite{HSNote,CelestePhD} \footnote{This is opposite to the $c=1$ case \cite{Das:1990kaa} where it is
the conjugate momentum that is rescaled in the change of variables
to (asymptotic) Liouville coordinates} 
\begin{eqnarray}
{\mathcal{H}}(\vec{\kappa})=\mathcal{H}\left(\vec{k},k^{z},\theta\right) & \equiv & \left.\frac{\eta_{\vec{k}_{1}\vec{k}_{2}}}{\frac{1}{|\vec{k}_{1}|}+\frac{1}{|\vec{k}_{2}|}}\right|_{\vec{k}_{1},\vec{k}_{2}\left(\vec{k},k^{z},\theta\right)}\label{Redefinition}\\
\Pi(\vec{\kappa})=\Pi_{\mathcal{H}}\left(\vec{k}.k^{z},\theta\right) & = & \left.\pi_{\vec{k}_{1}\vec{k}_{2}}\right|_{\vec{k}_{1},\vec{k}_{2}\left(\vec{k},k^{z},\theta\right)}.\label{MomRedefinition}
\end{eqnarray}

\noindent Since by construction (\ref{free energies}), $|\vec{k}_{1}|+|\vec{k}_{2}|=E=\sqrt{\vec{k}^{2}+(k^{z})^{2}}\equiv(P^{0})_{AdS}$,
the Hamiltonian can be written directly as an integral over $AdS\times S^{1}$:
\begin{eqnarray*}
H_{2} & = & \frac{1}{2}\int d\vec{k}_{AdS\times S^{1}}\hspace{3pt}\left[\left(\Pi_{\mathcal{H}}(\vec{k},k^{z},\theta)\Pi_{\mathcal{H}}(-\vec{k},-k^{z},-\theta)\right)\right.\\
 &  & +\left.\left(P^{0}\right)^{2}{\mathcal{H}}(\vec{k},k^{z},\theta){\mathcal{H}}(-\vec{k},-k^{z},-\theta)\right].
\end{eqnarray*}
In other words, we have obtained a bulk quadratic Hamiltonian in $AdS_{4}\times S^{1}$
by field redefinition and a simple change of variables: 
\begin{equation}
H_{2}=\frac{1}{2}\int d\vec{\kappa}\left[\Pi(\vec{\kappa})\Pi(-\vec{\kappa})+(P^{0})^{2}{\mathcal{H}}(\vec{\kappa}){\mathcal{H}}(-\vec{\kappa})\right].\label{BulkHam}
\end{equation}
We expand in spin fields \cite{Koch:2014aqa} 
\begin{equation}
h(k^{z},\vec{k},\theta)=\sum_{s=0, \pm2, ...}^{\infty} \frac{e^{i s\theta} }{\sqrt{\pi}} h_{s}(k^{z},\vec{k})\label{SpinExp}.
\end{equation}
The spin field $h_{s}(k^{z},\vec{k})$ can be further expanded \cite{Koch:2014aqa}, but for the purposes of this communication it is sufficient to observe the important property that $\theta \in [0,\pi]$. Indeed, from (\ref{map:4}), $\theta \sim \theta + \pi$, and the fact that $s$ is even follows.  This corresponds to the spectrum of the minimal
type A Vasiliev higher spin theory with a $\Delta=1$ scalar.

We then recognise (\ref{BulkHam}) as a sum of Hamiltonians
of massless spin $s$ fields in $AdS_{4}$\footnote{$\pi_{s}$ and $h_{s}$ are canonically conjugate fields}: 
\begin{equation}
H_{2}=\frac{1}{2}\sum_{s=0, \pm2, ...}^{\infty}\int d\vec{k}_{AdS}\hspace{3pt}[\pi_{s}(\vec{k},k^{z})\pi_{_{s}}(-\vec{k},-k^{z})+(P^{0})^{2}{h_{s}}(\vec{k},k^{z}){h_{s}}(-\vec{k},-k^{z})]\label{SpinSumHam}
\end{equation}

\subsection{Bulk higher spin fields at the IR critical point}


At the IR critical point, recall that the universal form \eqref{finite_scattering}
of the energy eigenstates was found to be: 
\begin{align*}
\eta_{\vec{k}_{1},\vec{k}_{2}}^{\vec{p}_{1},\vec{p}_{2}} & =\delta^{2}(\vec{p}_{1}-\vec{k}_{1})\delta^{2}(\vec{p}_{2}-\vec{k}_{2}) \\
 & +\frac{\left|p_{E}\right|}{\pi^{2}}\frac{\delta^{2}(\vec{p}_{1}+\vec{p}_{2}-\vec{k}_{1}-\vec{k}_{2})}{E_{\vec{p}_{1}\vec{p}_{2}}^{2}-\left(|\vec{k}_{1}|+|\vec{k}_{2}|\right)^{2}}\left(\frac{1}{|\vec{k}_{1}|}+\frac{1}{|\vec{k}_{2}|}\right).
\end{align*}
Further recall that $(\vec{k_{1}},\vec{k_{2}})$ are momentum coordinates
and that $(\vec{p_{1}},\vec{p_{2}})$ label the states with (squared)
energy $E_{\vec{p}_{1}\vec{p}_{2}}^{2}=(|\vec{p_{1}}|+|\vec{p}_{2}|)^{2}$.
We observe that 
\begin{align*}
|{p}_{E}| & =\sqrt{-\left(|p_{1}|+|p_{2}|\right)^{2}+\left(\vec{p}_{1}+\vec{p}_{2}\right)^{2}}\\
 & =i\sqrt{\left(|p_{1}|+|p_{2}|\right)^{2}-\left(\vec{p}_{1}+\vec{p}_{2}\right)\cdot\left(\vec{p}_{1}+\vec{p}_{2}\right)}\\
 & =i|p^{z}|.
\end{align*}
using the map (\ref{free energies}) - (\ref{map:4}).

The most general solution to the spectrum equations can then be written
as an arbitrary linear combination of the universal energy eigenstates:
\begin{eqnarray*}
\eta_{\vec{k}_{1}\vec{k}_{2}} & = & \int d\vec{p}_{1}\int d\vec{p}_{2}\,\,\eta_{\vec{k}_{1},\vec{k}_{2}}^{\vec{p}_{1},\vec{p}_{2}}\,\psi_{\vec{p}_{1}\vec{p}_{2}}\\
 & = & \psi_{\vec{k}_{1}\vec{k}_{2}}+\frac{i}{\pi^{2}}\int d\vec{p}_{1}\int d\vec{p}_{2}\left|p^{z}\right|\\
 &  & \frac{\delta^{d-1}(\vec{p}_{1}+\vec{p}_{2}-\vec{k}_{1}-\vec{k}_{2})\left(\frac{1}{|\vec{k}_{1}|}+\frac{1}{|\vec{k}_{2}|}\right)}{E_{\vec{p}_{1}\vec{p}_{2}}^{2}-\left(|\vec{k}_{1}|+|\vec{k}_{2}|\right)^{2}}\psi_{\vec{p}_{1}\vec{p}_{2}}.
\end{eqnarray*}
Following the prescription of the previous subsection, to change to
bulk higher spin variables, we make the identification 
\begin{eqnarray*}
H(\vec{\kappa}) & \equiv & \frac{\eta_{\vec{k}_{1}\vec{k}_{2}}}{\left(\frac{1}{|\vec{k}_{1}|}+\frac{1}{|\vec{k}_{2}|}\right)}\\
h(\vec{\kappa}) & \equiv & \frac{\psi_{\vec{k}_{1}\vec{k}_{2}}}{\left(\frac{1}{|\vec{k}_{1}|}+\frac{1}{|\vec{k}_{2}|}\right)},
\end{eqnarray*}
and change variables to $AdS_{4}\times S_{1}$ coordinates. Hence
we obtain, in the bulk: 
\begin{eqnarray*}
H(\vec{\kappa}) & = & h(\vec{\kappa})+\frac{i}{\pi^{2}}\int d\vec{p}\int dp^{z}\int d\theta\left|p^{z}\right|\\
 &  & \frac{\delta^{2}(\vec{p}-\vec{k})}{(p^{z})^{2}+(\vec{p})^{2}-(k^{z})^{2}-(\vec{k})^{2}}h(p^{z},\vec{p},\theta)\\
 & = & h(k^{z},\vec{k},\theta_{\vec{k}})\\
 &  & +\frac{i}{\pi^{2}}\int dp^{z}\int d\theta\frac{\left|p^{z}\right|}{(p^{z})^{2}-(k^{z})^{2}}h(p^{z},\vec{k},\theta).
\end{eqnarray*}
Under mild assumptions on the behaviour of $h(p^{z},\vec{k},\theta)$
as $|p^{z}|\rightarrow\infty$ (also requiring that $h(-k^{z})=h(k^{z})$),
one has 
\begin{eqnarray*}
\int dp^{z}\frac{\left|p^{z}\right|h(p^{z},\vec{k},\theta)}{(p^{z})^{2}-(k^{z})^{2}-i\epsilon}=i\pi h(k^{z},\vec{k},\theta),
\end{eqnarray*}
so that 
\begin{eqnarray}
H(k^{z},\vec{k},\theta) & = & h(k^{z},\vec{k},\theta)-\frac{1}{\pi}\int_{0}^{\pi}d\theta\hspace{3pt}h(k^{z},\vec{k},\theta).\label{Hs}
\end{eqnarray}
Expanding $h(k^{z},\vec{k},\theta)$ in spin $s$ fields as in (\ref{SpinExp}):
\begin{equation*}
h(k^{z},\vec{k},\theta)=\sum_{s=0, \pm2, ...}^{\infty} \frac{e^{i s\theta} }{\sqrt{\pi}} h_{s}(k^{z},\vec{k})
\end{equation*}
with $0<\theta<\pi$, we see that the latter term in equation (\ref{Hs})
precisely removes the $s=0$ field ($\Delta=s+1=1$) in the bulk,
and thus 
\begin{eqnarray*}
H(k^{z},\vec{k},\theta)=\sum_{s\neq0, \, s=\pm 2, ...}^{\infty}\frac{e^{i s\theta} }{\sqrt{\pi}} h_{s}(k^{z},\vec{k}),
\end{eqnarray*}
in agreement with \cite{Higher spin UIR}. 
More precisely, our result corresponds, in terms of higher spin representations
without $\Delta=1$, to the spectra of the (antisymmetric) direct
product of two $3d$ free $O(N)$ Majorana $\mathrm{Di}$ singletons
\footnote{The $SO\left(3,2\right)$ representations are labelled by $\left(\Delta,s\right)$
and $\mathrm{Di}=\left(1,\frac{1}{2}\right)$ }. That is, 
\begin{equation}
\left[\mathrm{Di}\otimes\mathrm{Di}\right]_{A}=\left(2,0\right)\varoplus\stackrel[s=1]{\infty}{\oplus}\left(2s+1,2s\right)
\end{equation}
The $3d$ free $O(N)$ Majorana fermion theory is dual to the minimal
type B Vasiliev higher spin theory which, in $d=3$, has the same spectra,
up to boundary conditions for the scalar field, as the minimal
type A Vasiliev higher spin theory with a $\Delta=2$ scalar \cite{Sezgin:2003pt,Leigh:2003gk} . This provides conclusive
evidence of the appropriateness of the identification of the bulk
higher spin fields as in (\ref{Redefinition}) and (\ref{MomRedefinition}).

It is left to observe that the interaction term $\frac{\lambda}{4!}\int d^{d-2}\vec{x}\left(\eta_{\vec{x}\vec{x}}\right)^{2}$
of the Hamiltonian does not contribute at the critical point, since
\begin{eqnarray*}
\eta_{\vec{x}\vec{x}}\sim1/\lambda+\mathcal{O}(1/\lambda^{2}).
\end{eqnarray*}
As such, the expression for the Hamiltonian (\ref{SpinSumHam}) simply
changes to exclude the $s=0$ term: 
\[
H_{2}=\frac{1}{2}\sum_{s\neq0, \, s=\pm 2, ...} \int d\vec{k}_{AdS}\hspace{3pt}[\pi_{s}(\vec{k},k^{z})\pi_{_{s}}(-\vec{k},-k^{z})+(P^{0})^{2}{h_{s}}(\vec{k},k^{z}){h_{s}}(-\vec{k},-k^{z})]
\]

\subsection{The $\Delta=2$ state and its Hamiltonian}

Returning to the $\Delta=2$ state, we recall that the particular
solution wave function of (\ref{eval_eqn}) is given by

\begin{align*}
\eta_{\vec{k}_{1}\vec{k}_{2}}^{B} & =\frac{\lambda}{48 \pi^2}\frac{1}{E^{2}-\left(|\vec{k}_{1}|+|\vec{k}_{2}|\right)^{2}}\left(\frac{1}{|\vec{k}_{1}|}+\frac{1}{|\vec{k}_{2}|}\right)J(\vec{k}_{1}+\vec{k}_{2}),\\
 & J(\vec{k}_{1}+\vec{k}_{2})=\int d^{2}l \, \eta_{\vec{k_{1}}+\vec{k_{2}}-\vec{l},\vec{l}}^{B}\,,
\end{align*}
provided 
\begin{equation}\label{BSEnerg}
E^{2}=E_{\vec{k}_{1}+\vec{k}_{2}}^{2}=(\vec{k}_{1}+\vec{k}_{2})^{2}-(\frac{\lambda}{48})^{2}.
\end{equation}
In other words, 
\[
\frac{\eta_{\vec{k}_{1}\vec{k}_{2}}^{B}}{\left(\frac{1}{|\vec{k}_{1}|}+\frac{1}{|\vec{k}_{2}|}\right)}=\frac{\lambda}{48 \pi^2}\frac{J(\vec{k}_{1}+\vec{k}_{2})}{(\vec{k}_{1}+\vec{k}_{2})^{2}-\left(|\vec{k}_{1}|+|\vec{k}_{2}|\right)^{2}-(\frac{\lambda}{48})^{2}}
\]
We can now implement the map (\ref{free energies}) - (\ref{map:4})
and the field redefinition (\ref{Redefinition}) to obtain the state
directly in the bulk: 
\begin{eqnarray}
H^{B}(\vec{\kappa})=-\frac{\lambda}{48 \pi^2}\,\frac{J(\vec{k})}{(k^{z})^{2}+(\frac{\lambda}{48})^{2}} ,\label{BState}\\
\int dk^{z}e^{ik^{z}z}H^{B}(\vec{\kappa})\sim e^{-|\lambda z|/48}.
\end{eqnarray}
The above bulk
description of the $\Delta=2$ state establishes it as a spin $0$
state with an exponential decay into the bulk. At criticality, the
state is then a boundary state, in agreement with \cite{Mulokwe:2018czu}. 

In order to obtain the Hamiltonian description of the state, we investigate the bulk properties of $J(\vec{k}_{1}+\vec{k}_{2})$:
\begin{align*}
J(\vec{k}_{1}+\vec{k}_{2})&= \int d^{2}l \, \eta_{\vec{k_{1}}+\vec{k_{2}}-\vec{l},\vec{l}}^{B} \\
&= \int d^{2}l_1 d^{2}l_2 \, \delta^2(\vec{k}_{1}+\vec{k}_{2}- \vec{l}_{1}+\vec{l}_{2}) \, \eta_{\vec{l}_{1}\vec{l}_{2}}^{B} \\
&= \int d\vec{l}_{AdS} \delta^2 (\vec{k} - \vec{l}) \, H^B (l^{z},\vec{l},\theta_{l})\\
&= \int d l^z \int d \theta_l  \, H^B (l^{z},\vec{k},\theta_l) = J(\vec{k}) \end{align*}
$J(\vec{k})$ is then a spin $0$ boundary ($z=0$) state. For finite $\lambda$, consistency of the solution (\ref{BState}) can be established directly in the bulk, requiring $\lambda < 0$.   

The interaction term in equation (\ref{quadham}) now takes the form: 
\begin{equation*}
\frac{\lambda}{4!}\int d^{2}\vec{x}\eta_{\vec{x}\vec{x}}^{2}= \frac{\lambda}{96 \pi^2} \int d^2k  J(\vec{k})   J(-\vec{k})    \end{equation*}
From the quadratic Hamiltonian (\ref{HS_ham1}) and the field redefinitions (\ref{Redefinition}) and (\ref{MomRedefinition}), it follows that $\Pi_{H}(\kappa)= \dot{H} (-\kappa)$, so that
\begin{equation}\label{ConjBound}
\Pi^B_{H}(\kappa)=-\frac{\lambda}{48 \pi^2}\,\frac{\Pi_J(-\vec{k})}{(k^{z})^{2}+(\frac{\lambda}{48})^{2}} , \end{equation}
where $\Pi_J (\vec{k})$ is the canonical conjugate to the boundary field $J(\vec{k})$. Substituting (\ref{BState}) and (\ref{ConjBound}) into the Hamiltonian (\ref{BulkHam}), performing the integrals over $k^z$ and (trivially) over $\theta $ and finally adding the above interaction contribution results in the Hamiltonian:
\begin{equation*}
H_2^B=\frac{24}{\pi^2 |\lambda|} \left\{ \frac{1}{2}  \int d^2k  \,   \Pi_J(\vec{k})\Pi_J(-\vec{k})+   \frac{1}{2}  \int d^2k  \, \left( \vec{k}^2- (\frac{\lambda}{48})^2  \right)J(\vec{k})J(-\vec{k})\right\}
\end{equation*}
This is, up to a factor, the expected Hamiltonian for a single mode with dispersion relation (\ref{BSEnerg}).

\section{Discussion and outlook}

In this paper, we built on the constructive approach which was developed
in Refs. \cite{Koch:2010cy,Koch:2014aqa} in both the light-cone gauge
and the temporal gauge for the free theory, in which an explicit map
between the conformal field theory in $d=2+1$ dimensions and the
higher spin theory in $AdS_{4}\times S_{1}$ was established. In the
Hamiltonian approach, the $1+2+2=5$ coordinates of the equal time
bilocals, map (in phase space) to the coordinates of AdS$_{4}\times S_{1}$.
We made use of the Hamiltonian approach in a time like gauge \cite{Koch:2014aqa},
and for the IR critical point, we considered an $O(N)$ vector theory
with a quartic interaction \cite{Mulokwe:2018czu}. The quartic interaction
contributes linearly in the bilocal field fluctuation equations, and
the spectrum problem is then that of a potential scattering problem.
The eigenstate solutions take a universal form at the critical point
\cite{Mulokwe:2018czu}.


The bulk description of these boundary eigenstates was obtained by
developing a remarkably simple first principles approach, consisting
of a simple change of variables from bilocal momenta to bulk momenta
(\ref{free energies}) - (\ref{map:4}), but requiring a field redefinition
in defining the bulk higher spin field. In this way, simple quadratic
Klein-Gordon bulk Hamiltonians are derived for the higher spin fields
in both UV and IR critical points, and, at the IR critical point,
the absence of an $AdS_{4}$ spin $0$ field is established directly
in the bulk. The $\Delta=2$ state is shown to be a boundary state
at IR criticality. Moreover, after integrating over $k^{z}$ and $\theta$, the boundary
quadratic Hamiltonian was obtained and has the expected dispersion
relation for the bound state. In future, it will be interesting to
look also at the non-decaying states which corresponds to mass deformations
in the bulk dual theory \cite{Aharony:2020omh}. 

The higher spin fields considered in this communication are different
from those in \cite{Koch:2014aqa}, but the approach should be equivalent,
at quadratic level, to the oscillator expressions obtained in that
article for the conformal generators. An extensive and comprehensive
study of the conformal algebra using the approach described in this
communication was carried out in \cite{CelestePhD} and shown to be
indeed equivalent to the bulk oscillator expressions obtained in \cite{Koch:2014aqa}.
This will be reported elsewhere \cite{JMR}.

It is of great interest to apply the approach developed in this communication
to generate interactions. The $1/N$ expansion of the collective field
Hamiltonian is well established (e.g.\cite{Demeterfi:1991cw} ,\cite{deMelloKoch:2012vc}).
Work in this direction is currently underway.




\end{document}